\documentclass[aps,twocolumn,longbibliography,pra,superscriptaddress,amssymb,amsmath,amsmath,floatfix]{revtex4-2}
\usepackage{graphicx}
\usepackage{epstopdf}
\usepackage{comment}
\usepackage{bm}
\usepackage{wrapfig}
\usepackage{times}
\usepackage[colorlinks=true,linkcolor=blue,urlcolor=blue,citecolor=blue]{hyperref}
\usepackage{float}
\usepackage{titlesec}
\usepackage{color,soul}

\begin{document}

%\title{Ultracold CaF and Ca interactions in ground and excited electronic states: implications to chemical reactions}
\title{Interactions and dynamics of metastable triplet-state helium dimer with Ne and Ar}

\author{Dibyendu Sardar}
\email{dibyendu.sardar@univ-rennes.fr}
\affiliation{Institut de Physique de Rennes, CNRS-Universit\'e de Rennes, UMR 6251, F-35000 Rennes, France}
%\affiliation{JILA, University of Colorado, Boulder, Colorado 80309, USA}

%\altaffiliation{Present address: Faculty of Physics, University of Warsaw, Poland}
\author{Fran\c{c}ois Lique}
\email{francois.lique@univ-rennes.fr}
\affiliation{Institut de Physique de Rennes, CNRS-Universit\'e de Rennes, UMR 6251, F-35000 Rennes, France}

\date{\today}% It is always \today, today,
\begin{abstract}
Helium dimer in its metastable triplet electronic state, He$_2(a~^3\Sigma_u^+)$, is a promising candidate for direct laser cooling and precision measurements. Interactions and collisions involving He$_2(a~^3\Sigma_u^+)$ remain unexplored in ultracold and cold energy domains.  Here, we present the first theoretical investigation of intermolecular interactions and collision dynamics of He$_2(a~^3\Sigma_u^+)$+Ne van der Waals complex. For the ground electronic state, $X~^3\mathrm{A}'$, we employ a spin-restricted coupled cluster method with single, double, and noniterative triple excitations. The resulting interaction potential exhibits an extremely shallow well with a depth of 4.612 cm$^{-1}$. We construct a three-dimensional vibrationally averaged potential energy surface. Using this surface, we perform full close-coupling calculations including fine-structure levels of He$_2(a~^3\Sigma_u^+)$ to investigate collisional excitation and quenching induced by Ne. We analyze the low-energy scattering dynamics and model resonant features associated with quasi-bound states supported by interaction potential. Propensity rule for fine-structure conserving  transitions is established. We additionally consider He$_2$+Ar collisions and compare their energy dependence and resonance structure with those of He$_2$+Ne. The He$_2$+Ar system exhibits a richer resonance structure, highlighting the sensitivity of low-energy scattering to interaction potential and reduced mass. Calculated PES and scattering properties provide a reliable benchmark for description of He$_2$+Ne van der Waals complex and may be useful for interpretation of future experimental spectra and collision measurements.

%Bound-state calculations reveal six bound levels of the He$_2$+Ne complex, from which we determine its spectroscopic properties.
%Calculated bound-state energies, spectroscopic constants, and scattering resonance characteristics provide a quantitative description of the He$_2$+Ne van der Waals complex and establish a theoretical basis for the interpretation of future experimental spectra and collision measurements.

%while  multireference configuration interaction method with single and double excitations is used for He$_2(^3\Sigma^+_u)$+Ar.while we construct a 2D surface is computed for He$_2(^3\Sigma^+_u)$+Ar under rigid rotor approximation for He$_2$.
\end{abstract}
\maketitle
%\begin{multicols}{2}
%----------------------------------------------------------------------------------------------
\section{Introduction}
\label{sec:intro}
%---------------------------------------------------------------------------------------------
In recent years, research on ultracold molecules (T $<1$ mK) has emerged as an impactful platform in atomic, molecular, and optical physics. Ultracold molecules have a range of applications from quantum computations \cite{DemillePRA2002,YelinPRA2006}, quantum simulations of many-body physics \cite{CornishNatPhys2024,MicheliNATPhys2006,BaranovCHEMREV2012}, precision measurements for fundamental physical constants \cite{CarrNJP2009,DemilleScience2017}, to high-resolution spectroscopy and ultracold chemistry \cite{LiuNAT2021,KarmanNP24}. Ultracold molecules are usually synthesized by association of pre-cooled atoms via magnetoassociation followed by an optical stabilization using stimulated Raman adiabatic passage \cite{VogesPRL2020,YangPRL2020,GuoPRL2016,ParkPRL2015,NiScience2008}. Another important method, direct laser cooling \cite{AndereggNATPHYS2018,CheukPRL2018,CaldwellPRL2019,DingPRX2020}, relies on repeated photon cycling enabled by a highly diagonal Franck-Condon factor between two involved electronic states. These experimental efforts are primarily focused on alkali \cite{VogesPRL2020,YangPRL2020,GuoPRL2016,ParkPRL2015,StevensonPRL2023} and alkaline-earth monofluorides \cite{AndereggNATPHYS2018,CheukPRL2018,CaldwellPRL2019} and oxides \cite{DingPRX2020}. This progress has enabled investigation of ultracold collision dynamics \cite{NiNature2010,GuoPRX2018}, resonances \cite{ParkNature2023}, and quantum state-resolved chemistry \cite{OspelkausScience2010,HuNatChem2021}. Alongside scattering cross-beam experiments \cite{SimonScience2013,JonghScience2020} have probed molecular collisions at energies approaching cold regime (T $<1$ K), revealed a rich spectrum of molecular resonances, and explored physical processes such as rotational transitions and vibrational relaxations. 

Yet, interactions and collision dynamics involving noble gas molecules remain largely unexplored in both ultracold and cold energy regimes. In this direction, metastable helium atoms have opened an initial platform. Bose-Einstein condensation of helium atoms in their metastable $2~^3S$ state has been achieved \cite{DosPRL2001}, followed by the realization of Fermi degeneracy \cite{McnamaraPRl2006} and determination of $s$-wave scattering length \cite{MoalPRL2006}. Additionally, collisions between two metastable helium atoms have been studied extensively \cite{VassenRMP2012}. Extending these achievements to He$_2$ molecules opens new opportunities in investigating fundamental physics and chemistry.

The major challenge is that conventional laser cooling techniques cannot be readily applied to cool He$_2$ molecule in its ground electronic state due to lack of the closed optical transitions. However, direct laser cooling may be feasible to the excited metastable state ($a~^3\Sigma_u^+$), which can have dipole-allowed transitions to high-lying levels. Recent experimental efforts have demonstrated prospects for direct laser cooling of He$_2$ dimer \cite{VerdegayIJP2025} in its metastable $a~^3\Sigma_u^+$ electronic state. This may open a novel route to realize noble-gas molecules at ultracold temperatures.  Metastable state of He$_2$ is intriguing and differs significantly from its ground $X ^1\Sigma_g^+$ electronic state. The ground $X ^1\Sigma_g^+$ state of helium dimer is very weakly bound and supports only a single vibrational state \cite{LuoJCP1992}. While excited metastable state supports multiple bound states, and spin conservation rule dictates its lifetime of 18 s \cite{ChabalowskiJCP1989}. Metastable He$_2$($a~^3\Sigma_u^+$) exhibits fine structure levels arising from spin-spin and spin-rotation interactions. High-precision spectroscopic measurements of He$_2$($a~^3\Sigma_u^+$) have recently been reported \cite{WirthPRA2025}. This system is currently the focus of active experimental efforts toward laser cooling and trapping.

%In ultracold and cold scattering experiments, helium is used as a coolant and a common choice of collisional partner. 
In this study, we focus on He$_2$($a~^3\Sigma_u^+$) dimer and investigate cold collisions with a noble-gas atom. He$_2$($a~^3\Sigma_u^+$) can be considered as a useful prototype system for studies of ultracold and cold collisions. The relatively larger energy spacing between rotational \cite{VerdegayIJP2025,WirthPRA2025} levels makes He$_2$($a~^3\Sigma_u^+$) well suited for quantum state-resolved scattering experiments. Nuclear-spin statistics theorem restricts allowed rotational levels of $^4$He$_2$($a~^3\Sigma_u^+$) to odd values of $N$, reducing number of rotational channels and thereby facilitating accurate quantum scattering calculations. Here we explore interactions and collisional properties of He$_2$($a~^3\Sigma_u^+$)+Ne and He$_2$($a~^3\Sigma_u^+$)+Ar van der Waals systems in their ground $X~^3\mathrm{A}'$ electronic state. While such calculations involving $X~^3\mathrm{A}''$ electronic state were extensively studied for a homonuclear O$_2$(X~$^3\Sigma_g^-$)+He \cite{LiqueJCP2010} and heteronuclear NH(X~$^3\Sigma_g^-$)+X (X = He, Ne, Ar) \cite{RamachandranJCP2018,NezaJCP2015,PrudenzanoJCP2019} systems at cold collision energies, both theoretically and experimentally. NH(X~$^3\Sigma_g^-$)+Mg collisions were investigated theoretically at ultracold temperatures \cite{GonzalezPRA2011}. However, interactions and collision properties involving molecular He$_2$($a~^3\Sigma_u^+$) remain unexplored theoretically and experimentally in both ultracold and cold collision energy regimes.  

To the best of our knowledge, this study, first theoretically investigate interactions and collisions of He$_2$($a~^3\Sigma_u^+$) with noble-gas atoms, considering Ne and Ar as colliding partners. Our primary objective is to develop highly accurate interaction potentials for these molecule-atom systems. We then perform quantum close-coupling scattering calculations to characterise and analyse collisional behaviour for a range of collision energies. These results could provide theoretical benchmarks to probe and guide future experimental studies of He$_2$($a~^3\Sigma_u^+$)+Ne collisions using crossed molecular-beam or spectroscopic experiments.

%Characterizing these interactions provides essential information on the collisional properties of metastable He$_2$ and establishes a theoretical basis for future experimental studies of He$_2$+noble-gas collisions.
%We first focus on He$_2$($a~^3\Sigma_u^+$)+Ne and investigate its bound-state structure and collision dynamics over a broad range of collision energies. We then compare the resulting scattering properties with those of the He$_2$($a~^3\Sigma_u^+$)+Ar system to examine the role of the noble-gas collision partner.

%For He$_2$($a~^3\Sigma_u^+$)+Ne, we determine the spectroscopic properties of the van der Waals complex and characterize its state-to-state collisional dynamics, including scattering resonances. 
We employ state-of-the-art electronic structure methods with large Gaussian basis sets to compute potential energy surfaces (PES) for the ground $X~^3$A$'$ electronic state of He$_2$($a~^3\Sigma_u^+$)+X (X = Ne, Ar). We use the coupled cluster method for He$_2$($a~^3\Sigma_u^+$)+Ne PES, while PES for He$_2$($a~^3\Sigma_u^+$)+Ar is computed using a configuration interaction method. We find that He$_2$($a~^3\Sigma_u^+$)+Ne van der Waals complex is very weakly bound and has less anisotropy in PES compared to He$_2$($a~^3\Sigma_u^+$)+Ar. For He$_2$($a~^3\Sigma_u^+$)+Ne complex, we construct a three-dimensional PES averaged over the ground vibrational state $v=0$ for He$_2$($a~^3\Sigma_u^+$). Exploiting this PES, we perform coupled-channel scattering and bound-state calculations explicitly including experimentally measured fine structure parameters of He$_2$($a~^3\Sigma_u^+$) \cite{VerdegayIJP2025,WirthPRA2025}. We compute state-to-state scattering cross sections for a wide range of collision energies. We observe scattering resonances at low collision energy, which are modelled and characterized by a partial wave analysis. For comparison, we also investigate collision dynamics of He$_2$($a~^3\Sigma_u^+$)+Ar. Notably, state-to-state scattering cross sections for He$_2$($a~^3\Sigma_u^+$)+Ar are larger in magnitude with richer resonance structure at low collision energies than those for He$_2$($a~^3\Sigma_u^+$)+Ne, highlighting the sensitivity of low-energy scattering to both reduced mass and interaction potential. We anticipate that our results provide a comprehensive characterization of He$_2$($a~^3\Sigma_u^+$)+Ne interaction and its low-energy collision dynamics, providing theoretical benchmarks for future experimental studies of metastable He$_2$+Ne collisions.

The structure of the paper is as follows. In Section \ref{sec:method}, we describe computational methods for electronic structure and collision dynamics. In Section \ref{sec:result}, we present and discuss results for interaction potential and collisional properties. Finally, in Section \ref{sec:summary}, we provide a summary and outlook of this work.

%----------------------------------------------------------------------------------------------
\section{Computations methods}
\label{sec:method}
\subsection{\textit{ab initio} calculation}
%----------------------------------------------------------------------------------------------
The interactions of a He$_2$ (a $^3\Sigma_u^+$) with ground-state of noble gas atoms, Ne, Ar, correlate with a $X~^3$A$'$ electronic state under C$_\mathrm{s}$ symmetry. We use state-of-the-art \textit{ab initio} quantum-chemical methods to compute PES for the ground-state of He$_2$+Ne and He$_2$+Ar van der Waals complexes. All \textit{ab initio} calculations are performed using the \textsc{Molpro} software package \cite{WernerJCP2020}. 

\begin{figure}[t]
  \includegraphics[width=0.38\textwidth]{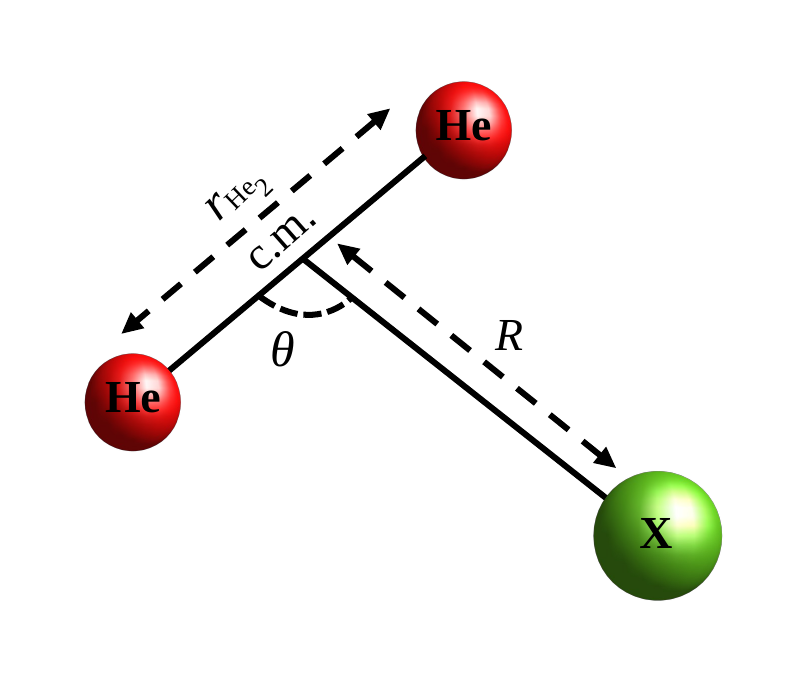}
  \caption{A schematic diagram of the He$_2$+X (X = Ne, Ar) system in the Jacobi coordinates. 
  }\label{fig:schematic}
\end{figure}

We present triatomic  He$_2$+Ne and He$_2$+Ar systems in Jacobi coordinates to compute two-dimensional (2D) and three-dimensional PESs. In Figure \ref{fig:schematic}, we present a schematic diagram for the Jacobi coordinates, where $R$ is the distance between the centre-of-mass (c.m.) of He$_2$ and X (X = Ne, Ar), and $\theta$ is the angle between the molecular axis and c.m. to X which defines rotation of He$_2$ dimer. For He$_2$+Ne van der Waals complex, we construct a vibrationally averaged 3D PES. We perform electronic structure calculations at five He$_2$ bond lengths, $r = [1.6, 1.8, 1.9894, 2.3, 2.6]$ bohr. This range of bond lengths allows us to account for the vibrational motion of the He$_2$ dimer up to $v=2$. For He$_2$+Ar system, we compute a 2D PES treating He$_2$ as a rigid rotor, and its bond length remains fixed at the equilibrium value, $r = 1.9894$ bohr \cite{DawidHe2dimer}. 

To accurately describe intermolecular interactions in He$_2$+X (X = Ne, Ar) complexes, we employ correlation-consistent quadrupole-$\zeta$ basis sets. In particular, we use aug-cc-pVQZ basis set \cite{KendallJCP1992,WoonJCP1993} for both Ne and Ar atoms. For He$_2$, we consider aug-cc-pCVQZ basis \cite{dawid} in our calculations. This new set of basis functions is particularly suitable for accurately describing the metastable helium dimer, in which one He atom is in the electronic ground state, and the other occupies the excited $2~^3S_1$ metastable state. These basis sets reproduce binding energy for the metastable $a~^3\Sigma_u^+$ electronic state of He$_2$ dimer with an accuracy of better than 10 cm$^{-1}$ \cite{DawidHe2dimer}. 

For the ground $X~^3$A$'$ electronic state of He$_2$+Ne complex, we employ highly correlated \textit{ab initio} electronic structure methods. We use a state-of-the-art partially spin-restricted coupled cluster method with single, double, and perturbative triple excitations, RCCSD(T). First, we construct converged partially open-shell Hartree-Fock (RHF) orbitals on a 2D grid of $R$ and $\theta$. These orbitals are then used as reference orbitals for subsequent RCCSD(T) calculations to determine the interaction energies over the entire $(r,R,\theta)$ grid. To ensure convergence, we perform \textit{ab initio} calculations in a distance loop starting from well-behaved RHF orbitals in the long range and use a wavefunction from that distance as an initial guess for next shorter distance. We consider 50 values of intermolecular distance $R$ in the range from 6 to 30 bohr. For the angular coordinate, we perform electronic-structure calculations at seven grid points, with $\theta$ ranging from $0^\circ$ to $90^\circ$ in steps of $15^\circ$. He$_2$ is a homonuclear diatomic molecule; the remaining angular range is related to this interval by symmetry and therefore does not require additional \textit{ab initio} calculations. We compute the interaction energy for He$_2$+Ne complex using the supermolecular approach, and we apply the Boys-Bernardi counterpoise correction for the basis-set superposition error.

We compute PES for ground $X~^3$A$'$ electronic state of He$_2$+Ar using the multireference configuration interaction method with single and double excitations (MRCISD). Our initial calculation reveals that  
the single-reference CCSD(T) method fails to converge at short intermolecular distances for linear ($\theta=0^\circ$) and quasi-linear ($\theta=15^\circ$) geometries. The $X~^3$A$'$ PES converges for all geometries in MRCISD method. To describe multireference character of He$_2$+Ar, we construct an active space (AS) composed of one core orbital, two singly occupied orbitals, and seven lowest-energy unoccupied orbitals for He$_2$. For Ar, we use 1s, 2s, 2p, 3s, 3p atomic orbitals in AS. We first perform a complete active space self-consistent field calculation on this AS, followed by the MRCISD calculation. Additionally, interaction energies are corrected for size consistency with the Davidson correction (MRCISD+Q).

The 3D PES for He$_2$+X is represented by an analytical functional form \cite{WernerJCP1989}:
\begin{equation}
    V(r,R,\theta)=\sum_{n=1}^{N_{\mathrm{max}}}\sum_{l=1}^{L_{\mathrm{max}}}d_{m,0}^{l+m-1}(\cos\theta)A_{nl}(R)(r-r_e)^{n-1},
    \label{eq:3DPES}
\end{equation}
where $d_{m,0}^{l+m-1}(\cos\theta)$ are Wigner rotation matrix elements, and $A_{nl}(R)$ is the radial expansion coefficients. $N_{\mathrm{max}}$ presents number of He$_2$ bond distances, $L_{\mathrm{max}}$ equals the number of angles $\theta$ for which He$_2$+X potential is computed,  $m=0$, and $r_e$ is the equilibrium bond distance for He$_2$. Under the rigid-rotor approximation for He$_2$, Eq.\ref{eq:3DPES} is expressed in terms of Legendre polynomials $P_\lambda(^.)$: 
\begin{equation}
    V(R,\theta) = \sum_{\lambda=0}^{\lambda_{\text{max}}} V_\lambda (R) P_\lambda(\cos\theta). \label{eq:2Dpes}
\end{equation}
Here, $\lambda=l-1$, and $V_\lambda(R)$ denotes the corresponding Legendre expansion coefficient of the 2D PES. The Legendre expansion coefficients provide a measure of the angular anisotropy of the interaction potential and are particularly useful in coupled-channel scattering calculations.

Averaging PES over vibrational wavefunction gives a more realistic interaction potential and generally provides better agreement with experiment than a simple 2D PES when pure rotational excitations are considered \cite{NezaJCP2015}. In the present work, we perform vibrational averaging only for He$_2$+Ne PES, thereby accounting for stretching of He$_2$ bond. The PES is averaged over vibrational states up to $v=2$ according to:
\begin{equation}
    V_v(R,\theta)=\langle v(r)\mid V(r,R,\theta\mid v'(r)\rangle,
    \label{eq:vibav}
\end{equation}
where $v(r)$ is the vibrational wave function for He$_2$, which is computed using the discrete variable representation method. He$_2$ potential used to obtain vibrational wavefunctions is calculated using the RCCSD(T) method with aug-cc-pCV6Z basis set. 

%In the present work, we use PES averaged over ground vibrational state $v=0$ for He$_2$ in He$_2$+Ne collision dynamics.

\begin{figure*}
\includegraphics[width=\linewidth]{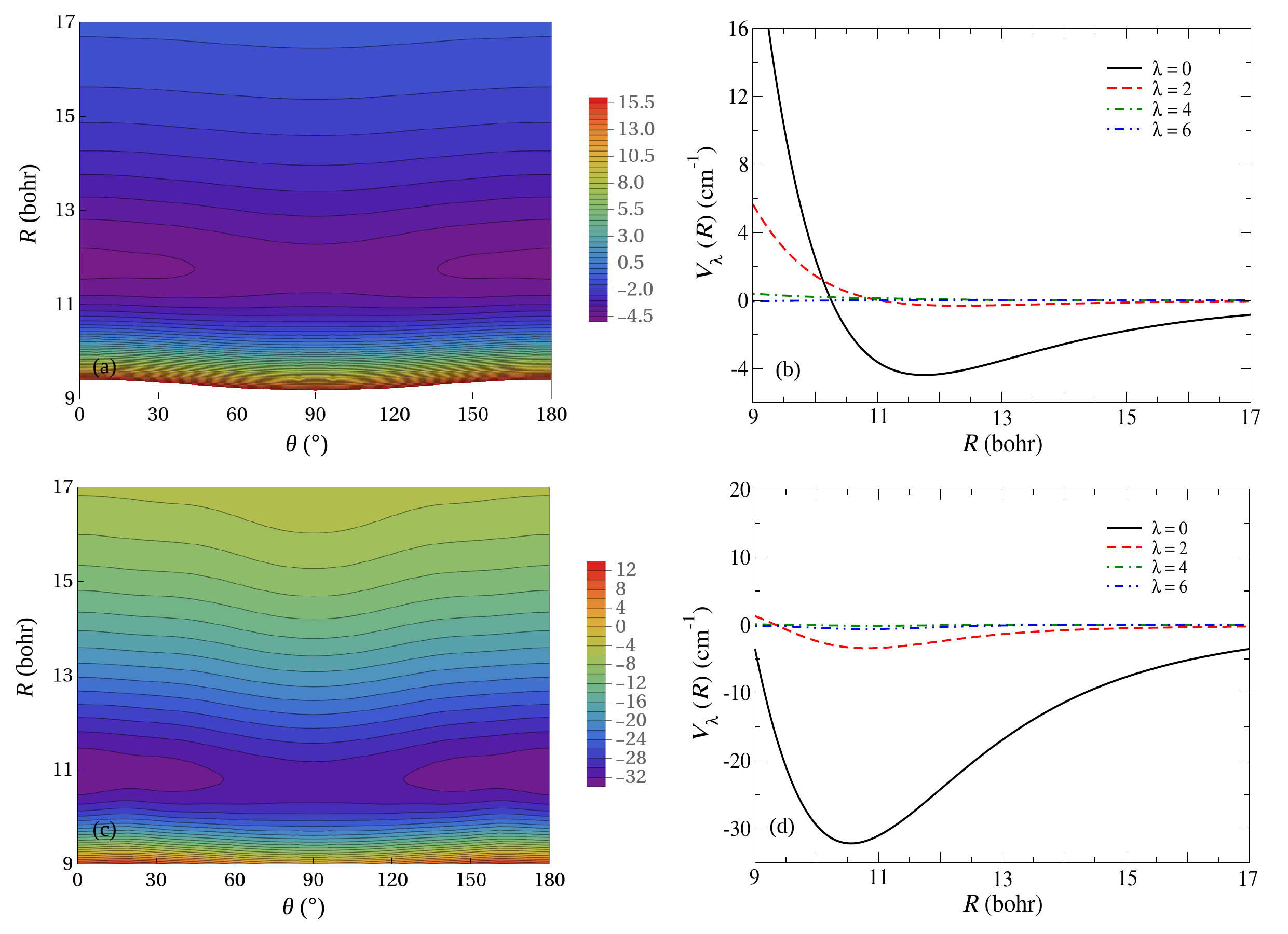}% 
\caption{\label{fig:legend}2D PESs (in cm$^{-1}$) and corresponding Legendre components for $X~^3\mathrm{A}'$ electronic state of He$_2$+Ne and He$_2$+Ar. Panels (a) and (b) show the 2D PESs and Legendre components for He$_2$+Ne, respectively, while panels (c) and (d) present the same for He$_2$+Ar.}
\end{figure*}

\subsection{Coupled-channel approach}
We carry out coupled-channel scattering and bound-state calculations on $X~^3\mathrm{A}'$ PES for He$_2$+Ne and He$_2$+Ar collision complexes. The Hamiltonian for these molecule+atom interacting pairs in the space-fixed (SF) coordinates is given by
\begin{equation}
    H=-\frac{\hbar^2}{2\mu R}\frac{\partial^2}{\partial R^2}R+ \frac{L^2}{2\mu R^2} + V(R, \theta) + H_{\mathrm{He}_2}, 
    \label{eq:totH}
\end{equation}
where first term is the relative kinetic energy of molecule+atom collision complex, and $\mu$ is the reduced mass. The second term is the centrifugal term that describes the end-over-end rotational energy of the interacting pair. The molecular Hamiltonian for He$_2$ on $a~^3\Sigma_u^+$ electronic state can be given by:
\begin{equation}
    H_{\mathrm{He}_2}= bN^2 + \frac{2}{3}\lambda_0(3S^2_\xi-S^2) + \gamma_0\boldsymbol{N.S},
    \label{eq:Hmol}
\end{equation}
where $b=7.589$ cm$^{-1}$ \cite{VerdegayIJP2025} is the rotational constant for He$_2$ in its $a~^3\Sigma_u^+$ electronic state. $\lambda_0=-0.036$ cm$^{-1}$ \cite{VerdegayIJP2025} and $\gamma_0=-0.0000807$ cm$^{-1}$ \cite{VerdegayIJP2025} are, respectively, spin-spin and spin-rotation coupling constants for $a~^3\Sigma_u^+$ state of He$_2$. $S_\xi$ is the component of electronic spin {\bf \textit{S}} on the He$_2$ molecular axis. The spin-spin coupling in molecular SF coordinates can be expressed in second-rank spherical tensor $\left[\boldsymbol{\mathit{S}} \otimes \boldsymbol{\mathit{S}}\right]^{(2)}_q$ as \cite{CybulskiJCP2005}:
\begin{equation}
\begin{split}
 \frac{2}{3}\lambda_0(3S^2_\xi-S^2)&=\frac{2}{3}\sqrt{6}\lambda_0\left[\frac{4\pi}{5}\right]^{1/2}\\ & \times \sum_q (-1)^qY_{2,-q}(\hat{\boldsymbol{\mathit{r}}})  \left[\boldsymbol{\mathit{S}} \otimes \boldsymbol{\mathit{S}}\right]^{(2)}_q ,
 \end{split}
\end{equation}
where $\hat{\boldsymbol{\mathit{r}}}$ denotes polar angles which describe direction of $\boldsymbol{\mathit{r}}$ in SF frame.

In this work, we solve coupled-channel equations subject to both scattering and bound-state boundary conditions. Scattering calculations are performed using the \textsc{Molscat} package \cite{HutsonCompPhy2019}. Such calculations yield scattering {\bf \textit{S}} matrix for a single value of total energy. Finally, we compute state-to-state scattering cross sections from the {\bf \textit{S}} matrix. In this work, we compute scattering cross sections for a wide range of energies, from ultracold regime ($10~ \mu$K) to the high-energy regime (350 K) with variable energy steps. In the range from $10~\mu$K to 10 K, we employ a dense energy grid to accurately characterise resonances. We include several energetically inaccessible levels, i.e., closed channels, to ensure convergence of calculated scattering cross sections. In this work, we use a He$_2$ rotational basis up to $N=9$ to ensure convergence of rotational cross sections between levels with $N\leq 3$. Additionally, we set a higher value of the total angular momentum quantum number until scattering cross sections converge with appropriately chosen convergence tolerances, DTOL and OTOL.

Coupled-channel bound states for He$_2$+Ne are obtained by using \textsc{Bound} package \cite{HutsonCompPhy1994}. This package uses the same rotational basis and interaction operators as \textsc{Molscat}. Bound states are calculated within the energy range $E_{\mathrm{min}}=-5$ cm$^{-1}$ and $E_{\mathrm{max}}= 2$ cm$^{-1}$ by solving coupled-channel equations. 

Both \textsc{Molscat} and \textsc{Bound} provide several different propagators for solving coupled-channel equations. In this work, coupled-channel equations are solved by using the Manolopoulos diabatic modified log-derivative propagator \cite{ManolopoulosJCP1986} for both scattering and bound states. We start propagation at $R_{\mathrm{min}}=6$ bohr for both scattering and bound-state calculations. For \textsc{Molscat} scattering calculations, we use $R_{\mathrm{max}}=2000$ bohr in the ultracold and cold energy regimes, while $R_{\mathrm{max}}=50$ bohr is used at relatively high collision energies. For the \textsc{Bound} calculations, we set the matching radius $R_{\mathrm{mid}}=10.5$ bohr where the calculation switches between short-range and long-range wave functions, and $R_{\mathrm{max}}=40$ with a propagator step size of 0.01 bohr.

\subsection{Basis sets}
For He$_2$ in $a~^3\Sigma_u^+$ electronic state, each rotational level $N$ is splitted into three fine structure components by spin-spin and spin-rotation interactions. In addition, we need to impose restrictions on the values of rotational angular momentum $N$ of He$_2 (a~^3\Sigma_u^+)$. Nuclear-spin statistics theorem eliminates all the even rotational levels for $a~^3\Sigma_u^+$ electronic state of He$_2$.

The eigen function of Eq \ref{eq:Hmol} for molecular Hamiltonian $H_{\mathrm{He_2}}$ in the intermediate coupling regime can be expressed as \cite{LiqueJCP2005,Gordy1984}:
\begin{equation}
\hspace{-0.2 in}
\begin{aligned}
\mid F_1jm\rangle &= \cos\alpha\mid N=j-1,Sjm\rangle
+ \sin\alpha\mid N=j+1,Smj\rangle,\\
\mid F_2jm\rangle &= \mid N=j,Sjm\rangle,\\
\mid F_3jm\rangle &= -\sin\alpha\mid N=j-1,Sjm\rangle
+ \cos\alpha\mid N=j+1,Smj\rangle,
\end{aligned}
\label{eq:F_states}
\end{equation}
where $\mid (N,S) jm\rangle$ presents basis functions for pure Hund's case (b). $\alpha$ is the mixing angle obtained by diagonalising molecular Hamiltonian, Eq.\ref{eq:Hmol}. Total molecular angular momentum $j$ corresponding to Hund's case (b) is given by
\begin{equation}
    \boldsymbol{\mathit{j}} = \boldsymbol{\mathit{N}} + \boldsymbol{\mathit{S}},
\end{equation}
where $\boldsymbol{\mathit{N}}$ and $\boldsymbol{\mathit{S}}$ are nuclear rotational and spin-angular momenta. In pure Hund's case (b), $\alpha\rightarrow0$, $F_1$ and $F_3$ levels correspond to $N=j-1$ and $N=j+1$, respectively.  For any value of $N$, fine structure energy levels for $a~^3\Sigma_u^+$ electronic state of He$_2$ follow the order: $F_3 > F_1 > F_2$.

%For the $a~^3\Sigma_u^+$ electronic state of He$_2$, spin-spin interaction is much stronger than spin-rotation, indicating largest contribution to the splitting between the fine structure levels comes from former interactions.

For the molecule+atom collision, molecular angular momentum $\boldsymbol{\mathit{j}}$ is coupled with orbital angular momentum $\boldsymbol{\mathit{L}}$ associated with the relative motion of the molecule and atom to form the total angular momentum $\boldsymbol{\mathit{J}}$ of the collision complex,
\begin{equation}
\boldsymbol{\mathit{J}}
=
\boldsymbol{\mathit{j}}
+
\boldsymbol{\mathit{L}},
\end{equation}
The coupled-channel basis is therefore represented by
\begin{equation}
\left|F_i,j,L;JM_J\right\rangle ,
\end{equation}
where $F_i$ labels the molecular fine-structure state, while $J$ and $M_J$ denote the total angular momentum and its SF projection, respectively. During collision process, conserved quantum numbers are $J$, $M_J$ and total parity $p=(-1)^{N+L+1}$. For a given $j$ and $J$, the allowed values of $L$ satisfy a triangular relation: $\mid j-J\mid \le L \le \mid j+J\mid$ subject to the parity requirement.

%----------------------------------------------------------------------------------------------
\section{Results and discussions}
\label{sec:result}
\subsection{PES}
%----------------------------------------------------------------------------------------------

Panel (a) of Figure \ref{fig:legend} presents a 2D contour plot of the 3D PES for $X~^3\mathrm{A}'$ electronic state of the He$_2$+Ne complex. The PES is obtained using the RCCSD(T) method and averaged over the ground vibrational state ($v=0$) of He$_2$. Panel (c) presents corresponding 2D PES for He$_2$+Ar complex, calculated using MRCISD+Q method. The overall shapes of the two PESs are very similar. We note that for both He$_2$+Ne and He$_2$+Ar complexes, the analytic potential reproduces the calculated interaction energies quite well. Over the entire grid, we checked that the mean relative difference between the analytic fit and the computed \textit{ab initio} interaction energies is less than 1.5\%. The first four nonzero even-order Legendre components of the respective PESs are shown in panels (b) and (d), respectively. Notably, He$_2$+Ne/Ar PES is symmetric with respect to the exchange of the two identical He atoms, corresponding to $\theta\rightarrow\pi-\theta$. Consequently, all odd-order Legendre components vanish, and only even values of $\lambda$ contribute to the PES expansion. For He$_2$+Ar, the $V_2$ and $V_4$ components are larger and contribute more significantly relative to the isotropic component $V_0$ than those for the He$_2$+Ne system. Therefore, the He$_2$+Ar PES exhibits stronger angular anisotropy than the He$_2$+Ne PES. Consequences of anisotropy of interaction potentials for scattering cross sections are discussed in the following sections. 

%Notably, owing to the homonuclear symmetry of He$_2$, all odd-order Legendre components vanish, and only even values of $\lambda$ contribute to the PES expansion.

For the He$_2$+Ne complex, we determine global equilibrium geometries from the analytic PES. The global minimum for He$_2$+Ne complex occurs at a linear geometry, with $\theta=0^\circ$ and $R=11.170$ bohr, and corresponds to an extremely shallow potential well of approximately 4.612 cm$^{-1}$. In addition, we also performed full-dimensional geometry optimizations at the RCCSD(T) level to determine the global minimum for ground $X~^3\mathrm{A}'$ electronic state of He$_2$+Ne.  A similar equilibrium properties are reproduced as in analytic PES of He$_2$+Ne. In comparison, the potential well depth of the He$_2$+Ar complex is 33.518 cm$^{-1}$. The substantially deeper potential well of He$_2$+Ar can be attributed to higher polarizability of Ar than Ne, resulting in stronger dispersion interactions. Interaction strength of He$_2$+Ne is also considerably weaker than that of the ground $^3\mathrm{A}''$ state of prototype cold molecule NH in NH+X (X = Ne, Ar) van der Waals complexes \cite{RamachandranJCP2018,NezaJCP2015,PrudenzanoJCP2019}. In particular, the well depth of He$_2$+Ne is smaller than that of NH+Ne, whereas the well depth of He$_2$+Ar is comparable to that of NH+Ne \cite{NezaJCP2015} and smaller than that of NH+Ar \cite{PrudenzanoJCP2019}.

\begin{table}
\caption{\label{tab:bound} He$_2$+Ne bound energy levels in cm$^{-1}$ with inclusion of He$_2$ fine structure.}
\begin{ruledtabular}
\begin{tabular}{ccc}
% &\multicolumn{2}{c}{global geometry}&\multicolumn{2}{c}{local geometry}\\
 $J$  & $L$ & Energy (cm$^{-1}$)  \\
\hline
 1  & 0 & -1.832  \\ 
 2  & 0 & -1.794  \\ 
 0  & 0 & -1.757  \\ 
 1  & 1 & -1.736  \\ 
 2  & 1 & -1.701  \\ 
 0  & 1 & -1.694  \\ 
 3  & 1 & -1.681  \\ 
 1  & 1 & -1.675  \\ 
 2  & 1 & -1.646  \\ 
 1  & 1 & -1.616 \\
 2  & 2 & -1.469  \\
 \end{tabular}
\end{ruledtabular}
\end{table}

\begin{table}
\caption{\label{tab:bound_nofine} He$_2$+Ne bound energy levels in cm$^{-1}$ without He$_2$ fine structure.}
\begin{ruledtabular}
\begin{tabular}{ccc}
% &\multicolumn{2}{c}{global geometry}&\multicolumn{2}{c}{local geometry}\\
 $J$  &  $L$ & Energy (cm$^{-1}$)  \\
\hline
 1 &  0 & -1.828  \\ 
 0 &  1 & -1.761  \\ 
 2 &  1 & -1.706  \\ 
 1 &  1 & -1.672  \\ 
 2 & 2  & -1.443  \\ 

 \end{tabular}
\end{ruledtabular}
\end{table}

\begin{figure}[t]
  \includegraphics[width=\linewidth]{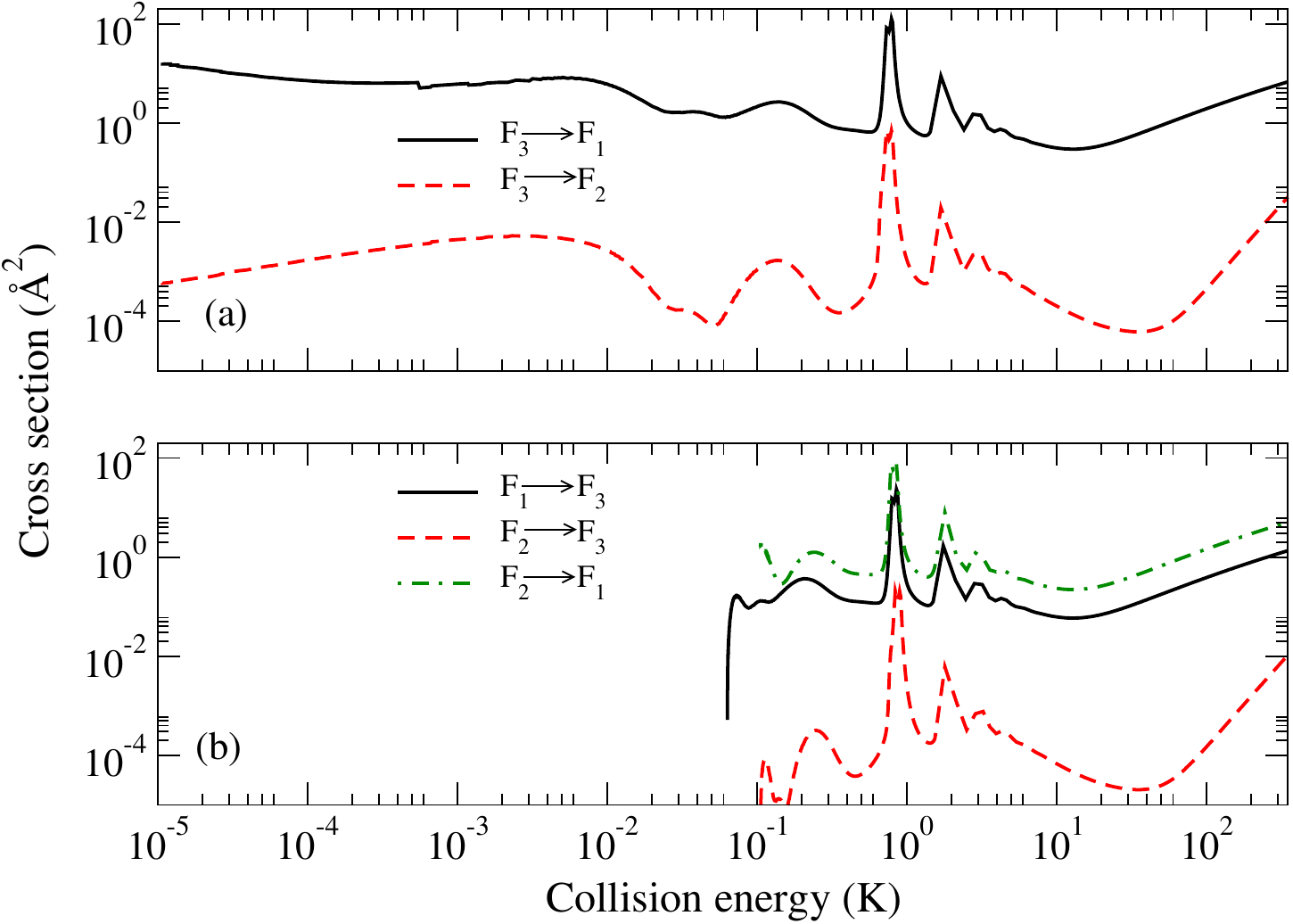}
  \caption{ He$_2$+Ne scattering cross sections (in $\mathrm{\AA}^2$) for quenching or de-excitation and excitation transitions within $N=1$ are shown in Panel (a) and (b), respectively. Both the cross-section and collision energy axes are on a logarithmic scale. 
  }\label{fig:N1block}
\end{figure} 

\subsection{Bound states}
We calculate bound-state energy levels for He$_2$+Ne van der Waals complex using the 3D PES averaged over ground vibrational state $v=0$; hereafter, we call it 3D-ave-PES. Bound state calculations are performed for the $^4$He and $^{20}$Ne isotopes employing the coupled-channel approach as implemented in \textsc{Bound} package. He$_2$ molecule in its $a~^3\Sigma_u^+$ electronic state exhibits fine structure due to coupling between rotational momentum and electronic spin. We modify the \textsc{Bound} programme to explicitly include fine structure of He$_2$. Rotational basis we include for the rotational states with quantum number $N\le9$.

In Table \ref{tab:bound}, we present bound-state energies for the first few values of total angular momentum $J$ of He$_2$+Ne complex computed with the 3D-ave-PES. Computed bound state energies are given relative to the fine structure $F_2$ of He$_2$($a~^3\Sigma_u^+$). The lowest bound state is characterised by $L=0$ with odd parity, and total angular momentum $J=1$. We also calculate bound-state energies for He$_2$+Ne complex, excluding fine structure of He$_2$. The basis set and propagation parameters remain the same for this calculation. The resulting energy levels are listed in Table \ref{tab:bound_nofine}. The bound-state energies calculated with and without fine-structure coupling are in close agreement. This indicates that fine-structure coupling has only a weak effect on the binding energies of the He$_2$+Ne complex, a similar behaviour reported for NH+He \cite{RamachandranJCP2018}. Since, He$_2$+Ne complex has a linear global equilibrium geometry, we fit bound-state energies listed in Table \ref{tab:bound_nofine} to the rigid rotor energy expression:
\begin{equation}
    E_J = E_0 + BJ(J+1)-DJ^2(J+1)^2.
    \label{eq:rigidrotor}
\end{equation}
Here, $E_0$ is the energy of the $J=0$ level, while $B$ and $D$ are the rotational and quartic centrifugal-distortion constants, respectively, for the He$_2$+Ne van der Waals complex.  Using the least-squares method, we obtain $B=0.053$ cm$^{-1}$, and $D=0.003999$ cm$^{-1}$. 

The calculated bound-state energies could be measured in future crossed molecular-beam or spectroscopic experiments, providing an opportunity to test the predicted interaction potential of He$_2$+Ne complex. Additionally, calculated spectroscopic constants could provide a reliable description of the rotational energy structure of the He$_2$+Ne van der Waals complex and are useful for the interpretation of future experimental spectra.

%Moreover, the 3D-ave-PES could provide a useful theoretical framework for interpreting resonant features that may be observed in future crossed molecular-beam measurements of collision cross sections.

\subsection{Collision dynamics}
\subsubsection{He$_2$+Ne}

\begin{figure}[t]
  \includegraphics[width=\linewidth]{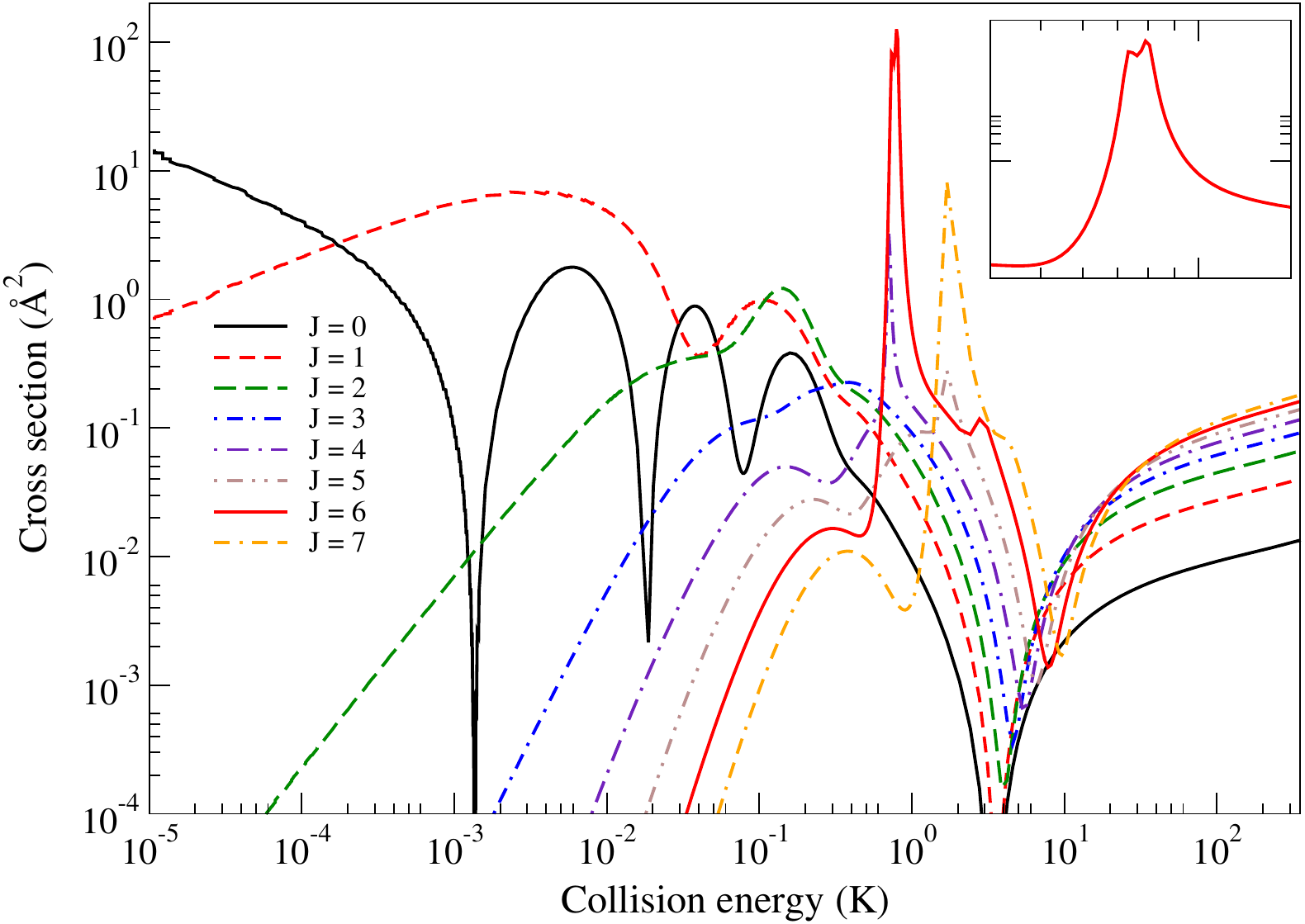}
  \caption{Scattering cross sections for $F_3 \rightarrow F_1$ transition for different values of $J$. The sub-panel shows an enlarged view of the resonant peaks associated with $J=6$ state. 
  }\label{fig:res}
\end{figure} 

We study collisional excitation and rotational quenching of He$_2$ by Ne using the 3D-ave-PES. We use main $^4$He and $^{20}$Ne isotopes in scattering calculations. The quantal coupled-channel equations for He$_2$+Ne scattering are solved in the intermediate coupling scheme using the \textsc{Molscat} package, modified to take into account fine structure of rotational energy levels. The reduced mass for He$_2$+Ne system is $\mu=5.716$ amu.

In Figure \ref{fig:N1block}, we present state-to-state inelastic scattering cross sections as a function of collision energy for quenching and excitation transitions within $N=1$ rotational level for He$_2$. We note that cross sections for both transitions converge at total angular momentum $J=23$ for a collision energy of 10 K. We observe that there is a significant contribution from quenching cross section for $F_3\rightarrow F_1$ transition in the ultracold energies, while such cross section for $F_3\rightarrow F_2$ transition is very small and approaches zero.  In the limit of ultracold energy, He$_2$+Ne collisions are dominated by $s$-wave scattering ($J=0$), and cross section for $F_3\rightarrow F_1$ transition follows the Wigner threshold law, varying approximately as the inverse of the collision velocity \cite{CybulskiJCP2005}. However, the $F_3\rightarrow F_2$ transition is strongly suppressed at low collision energies and the corresponding cross section approaches zero. Such suppression is a consequence of a forbidden or indirectly coupled transition, as discussed for collisions in C+He \cite{StaemmlerJPB1991} and O$_2$+H$_2$ \cite{SimonScience2013} systems. In the collision energy regimes of 1-350 K, the variations of cross sections for quenching and excitation transitions exhibit similar energy dependence. We observe resonant peaks in cross sections near 0.75 K for transitions between fine structure levels. 

%These resonant features may arise from either shape or Feshbach resonances.

To investigate the origin of two-peak structure in scattering cross sections near 0.75 K, we perform a partial wave analysis \cite{PlompJPCL2024}. In particular, we examine partial wave contribution of each relevant total angular momentum state $J$ to total scattering cross section for a particular transition. For this, we consider $F_3\rightarrow F_1$ quenching transition within $N=1$ rotational level. In Figure \ref{fig:res}, we show de-excitation transition for $J=0-7$ as a function of collision energy. In the sub-panel of Figure \ref{fig:res}, we show a resonant two-peak structure for $J=6$ state. The rapid variation of cross sections with energy for individual $J$  with distinct peaks and valleys may correspond to scattering resonances. We find that a two-peak structure near collision energy 0.75 K arises predominantly from the contribution of $J=6$ total angular momentum state. 

\begin{figure}[t]
  \includegraphics[width=\linewidth]{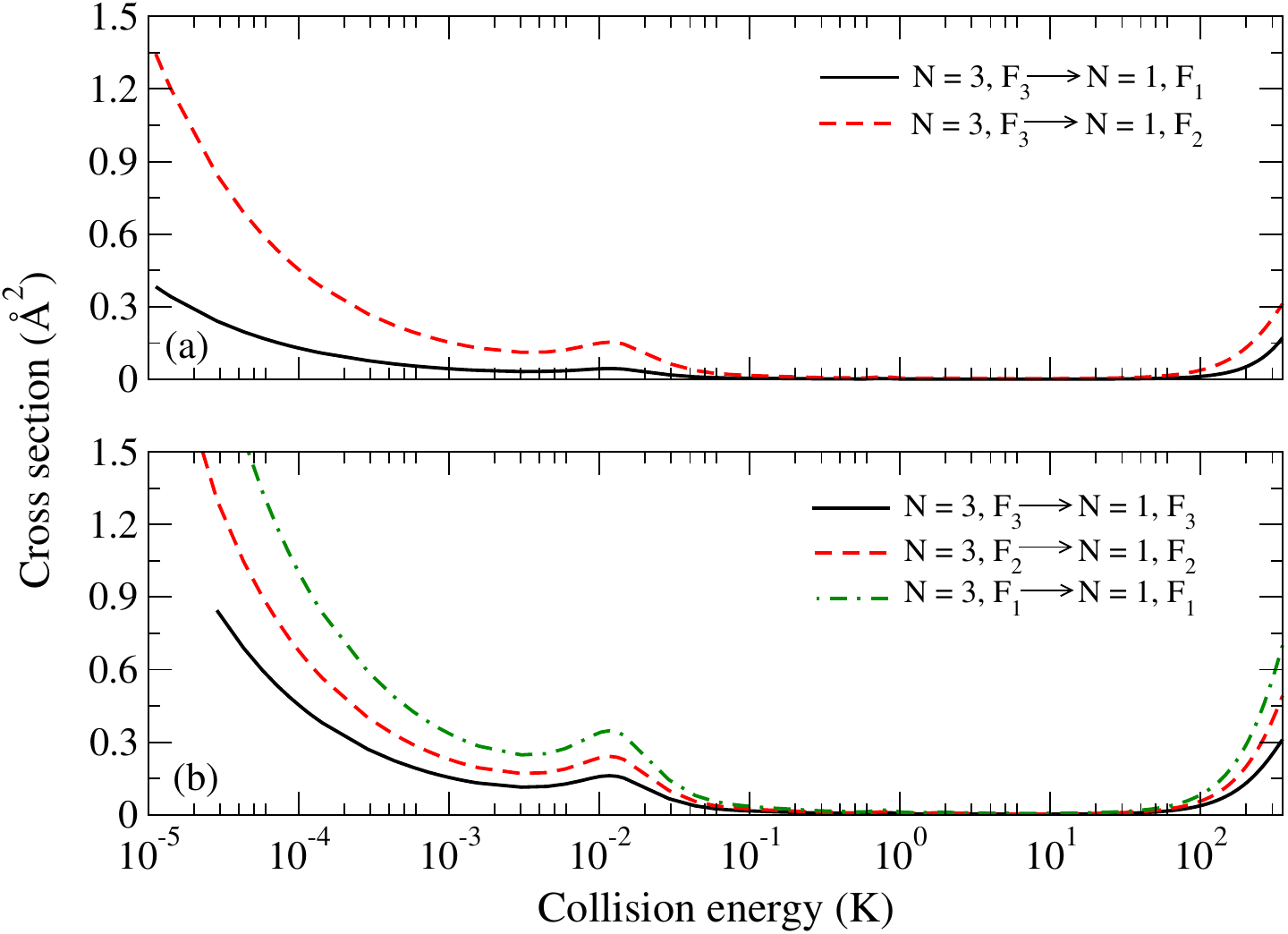}
  \caption{Scattering cross sections for fine structure changing and fine structure conserving transitions in He$_2$+Ne collisions are in Panel (a), and (b), respectively. 
  }\label{fig:quench}
\end{figure} 

To further investigate the resonant origin, we calculate bound states of He$_2$+Ne collision complex for $J=6$ over the energy range $E_{\textrm{min}}=-7.5$ K to $E_{\textrm{max}}=3$ K. We identify two quasi-bound states with energies close to 0.85 K when the internal energy of the initial $F_3$ level, $E_{F_3}=0.105$ K, is included. The centrifugal barrier associated with the $J=6$ partial wave can temporarily trap the collision complex within the attractive region of the He$_2$+Ne interaction potential, giving rise to quasi-bound states and the associated resonant features. The close correspondence between the energies of these quasi-bound states and the observed scattering peaks supports their assignment as shape resonances. Similar shape resonances have been reported at low collisional energies for H$_2$+O$_2$ collisions \cite{SimonScience2013}.

For He$_2$+Ne collisions, only even $\Delta N$ transitions will be observed in the rotational spectrum, while transitions with Odd $\Delta N$ will be missing. This is a consequence of the presence of only even-order anisotropic Legendre components of PES.
In Figure \ref{fig:quench}, we show cross sections for rotational quenching transitions from $N=3\rightarrow N=1$ within different fine-structure levels as a function of collision energy. Panel (a) presents cross sections for fine-structure changing transitions, while cross sections for fine-structure conserving transitions are shown in panel (b). In ultracold energy limit, cross sections increase for both transitions due to the Wigner threshold effect. In the energy limit 1-350 K, magnitude of cross sections shown in Figure \ref{fig:quench} seems to be governed by following propensity rule:

\begin{figure}
  \includegraphics[width=\linewidth]{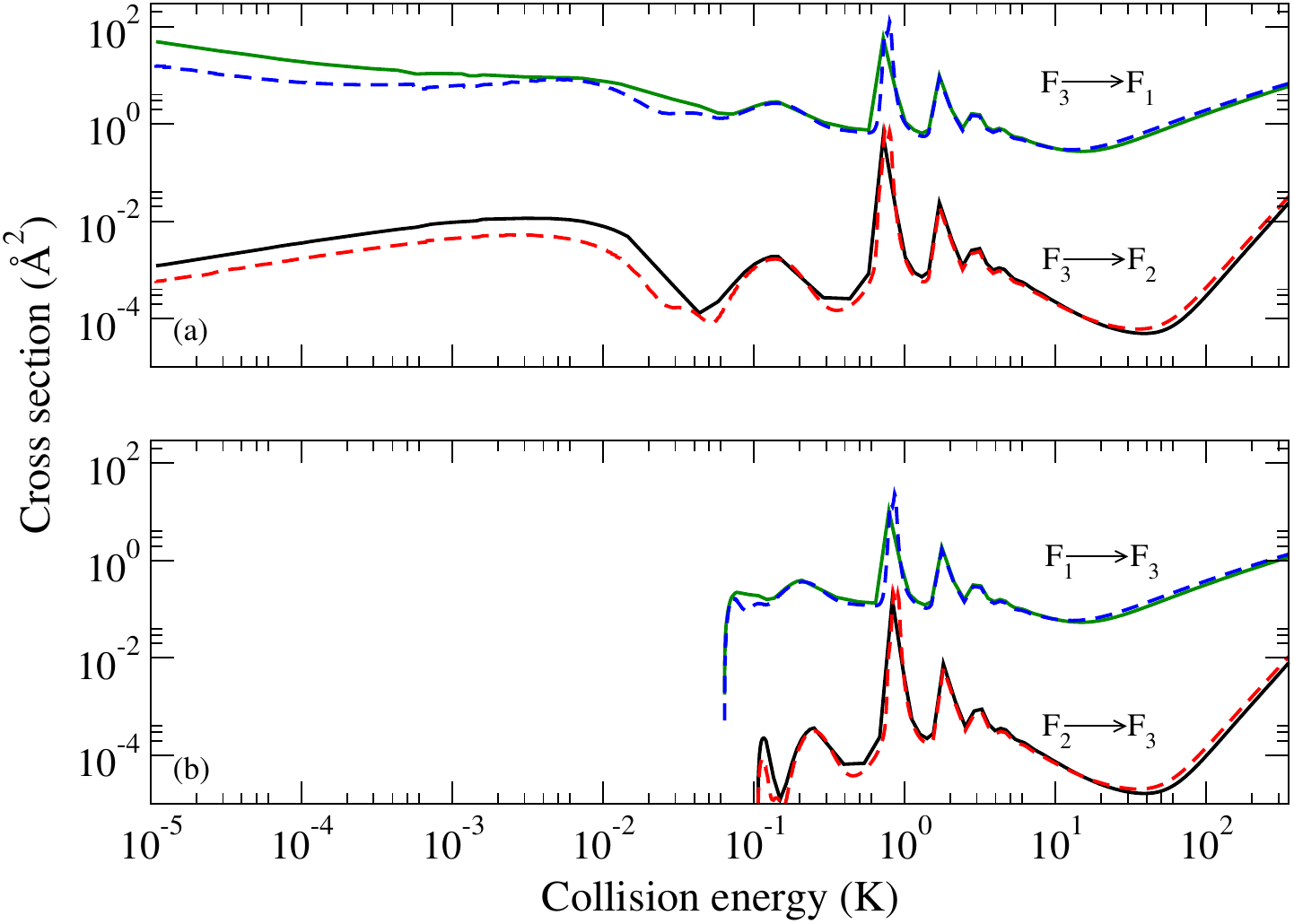}
  \caption{Scattering cross sections as a function of collision energy for quenching and excitation transitions in He$_2$+Ne collisions are shown in Panel (a) and (b), respectively, calculated using 3D-ave-pes (dashed curves) and 2D PES (solid curves). 
  }\label{fig:2Dvs3D}
\end{figure}

\begin{itemize}
      \item A propensity rule exists for $F_i$ conserving transitions [$\Delta J = \Delta N$ for pure Hund's case (b)]. In the present He$_2$+Ne system, the cross sections for fine-structure conserving transitions are slightly larger than those for fine-structure changing transitions.
\end{itemize}
The propensity rule where fine-structure conserving transition are more favorable over fine-structure changing transition and is general for molecules in $^3\Sigma^-_g$  electronic state and discussed in details for NH($^3\Sigma^-_g$)+He \cite{RamachandranJCP2018}, O$_2$($^3\Sigma^-_g$)+He \cite{LiqueJCP2010}, and SO($^3\Sigma^-_g$)+He \cite{LiqueJCP2005} collisional systems.

\subsubsection{2D vs 3D averaged PES}

In addition to calculations of He$_2$+Ne scattering dynamics using 3D-ave-PES, we also compute scattering cross sections using He$_2$+Ne 2D PES. In 2D PES, He$_2$ intermolecular separation is fixed at its equilibrium value. In Figure \ref{fig:2Dvs3D}, we compare scattering cross sections for quenching and excitation transitions within the ground rotational manifold obtained using the two PESs. Solid curves correspond to cross sections calculated using 2D PES, while dashed curves represent those using 3D-ave-PES. For both PESs, cross sections exhibit a similar energy dependence, however they differ in magnitude. The difference is significant in the ultracold regime, and becomes smaller at intermediate and higher collision energies.

We check that at ultracold collision energy, the cross section for $F_3 \rightarrow F_1$ transition differs by approximately 66\% between two PESs. The corresponding difference can be as large as 7\% -- 20\% for low collision energies, where percentage difference is calculated relative to the cross section obtained using the 2D PES. Despite these quantitative differences, the 2D PES yields qualitatively similar He$_2$+Ne scattering properties to those obtained with the 3D-ave-PES. The overall shapes of scattering cross sections and position of resonances for both quenching and excitation transitions are nearly the same for two PESs. Previous studies \cite{RamachandranJCP2018,NezaJCP2015} have shown that vibrationally averaged PES reproduces the experimental results more accurately than a 2D PES. Thus, present comparison suggests that vibrational averaging seems to be important for obtaining quantitatively accurate cross sections that can be compared with future experimental measurements.

\subsubsection{He$_2$+Ar}
\begin{figure}
  \includegraphics[width=\linewidth]{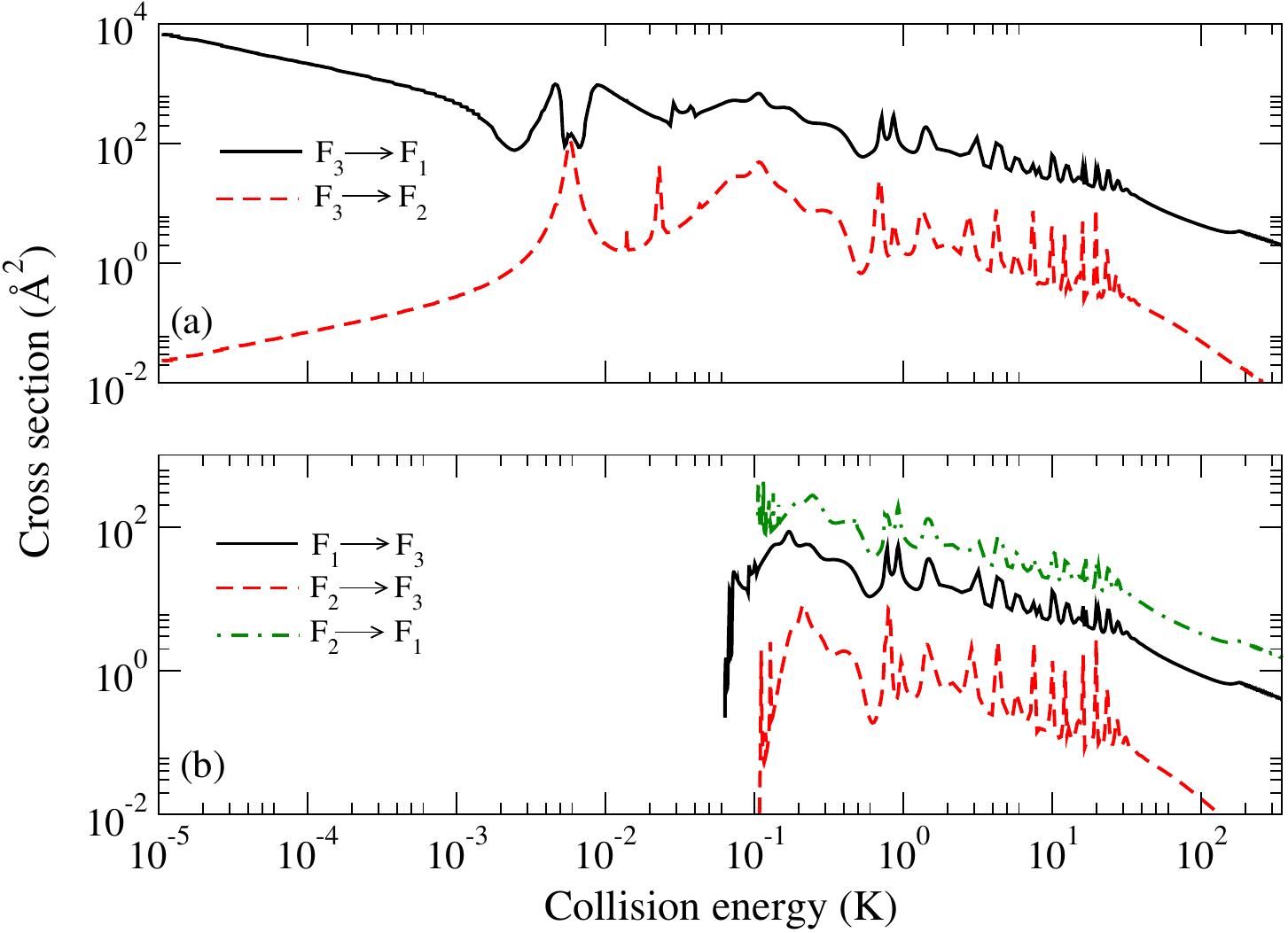}
  \caption{Scattering cross sections as a function of collision energy for He$_2$+Ar collision system. Panel (a) presents quenching cross sections, while excitation cross sections are in Panel (b). 
  }\label{fig:He2+Ar}
\end{figure}

Having established the effect of dimensionality and vibrational averaging for the He$_2$+Ne system, we also study He$_2$+Ar collisions over a range of collision energies. We compare the energy dependence and resonance features of the scattering cross sections of He$_2$+Ar with those of He$_2$+Ne system. For this purpose, we employ He$_2$+Ar 2D interaction potential and calculate fine structure changing cross sections within the ground rotational manifold. The aim is not to provide a detailed analysis of the He$_2$+Ar scattering dynamics, but rather to examine the similarities and differences in the energy dependence of the cross sections and the associated resonance structures. In Figure \ref{fig:He2+Ar}, we present the calculated scattering cross sections as a function of collision energy. In the ultracold regime, the quenching cross section for $F_3\rightarrow F_1$ transition increases with decreasing collision energy, consistent with the Wigner threshold behaviour. At higher collision energies, several peaks are observed, indicating the presence of scattering resonances. 

These resonances can arise from quasi-bound states supported by the attractive He$_2$+Ar interaction potential in combination with the centrifugal barrier, which temporarily confines the collision complex before dissociation. The attractive well of the He$_2$+Ar potential has a depth of approximately 33.518 cm$^{-1}$, substantially larger than that of the He$_2$+Ne system, and supports a richer resonance structure. A detailed assignment of the individual resonances is beyond the scope of the present paper. Overall, the energy dependence of the He$_2$+Ar cross sections differs from that of He$_2$+Ne, with a larger number of resonant features in former system. The deeper interaction potential, together with differences in the reduced mass and anisotropy, may contribute to the richer resonance structure observed for He$_2$+Ar.

%==============================================================================================
\section{Conclusions}
\label{sec:summary}
We have investigated intermolecular interactions and collision dynamics between a metastable triplet state He$_2$ molecule and a noble gas atom, Ne, as the collider. Using state-of-the-art \textit{ab initio} quantum chemistry methods, we computed PES for the ground $X~^3\mathrm{A'}$ electronic state of He$_2$($^3\Sigma_u^+$)+Ne. By averaging over ground vibrational wave function of He$_2$, we constructed an accurate 3D vibrationally averaged interaction potential for He$_2$($^3\Sigma_u^+$)+Ne. The resulting potential exhibits weak anisotropic character. Notably, He$_2$+Ne van der Waals complex is extremely weakly bound, with a binding energy of 4.612 cm$^{-1}$.

By using 3D-vib-ave PES, we performed quantum coupled-channel bound-state and scattering calculations 
for He$_2$+Ne collision complex. In coupled-channel calculations, we considered fine structure for He$_2$ dimer. The computed lowest bound state energy for He$_2$+Ne complex is -1.832 cm$^{-1}$, characterised by $L=0$ within parity block 2. We computed rotational constant $B=0.0627$ cm$^{-1}$ and quartic centrifugal distortion constant $D=0.000129$ cm$^{-1}$ for He$_2$+Ne. We further calculated scattering cross sections for quenching and excitation transitions within a wide range of collision energies.  Resonant features were observed in cross sections at low collision energy. We modelled and analysed resonant peaks by partial wave analysis. The centrifugal barrier associated with $J=6$ can temporarily trap collision complex within attractive part of He$_2$+Ne potential supports two quasi-bound states, resulting in observed resonances. 

For comparison, we also constructed a 2D interaction potential and investigated collisional dynamics of He$_2$+Ar. The He$_2$+Ar system has richer resonant structure in scattering cross sections than He$_2$+Ne. The deeper potential well, stronger anisotropy, and higher reduced mass may be attributed to observed resonances in He$_2$+Ar. The first vibrationally averaged PES reported for He$_2$+Ne complex, together with the calculated bound-state and collision-dynamics data, provides an initial theoretical testbed for future experimental studies. These results may be useful for interpreting future crossed-molecular-beam and spectroscopic measurements of the He$_2$+Ne complex.

%---------------------------------------------------------------------------------------------

\begin{acknowledgments}
We are thankful to Jacek K{\l}os and Marcin Gronowski for discussions in computing PES.
We gratefully acknowledge support of the European Union through the Marie Sk{\l}odowska-Curie Action (MSCA)-Postdoctoral Fellowship HighSpin$\_$Helium, under Grant Agreement No. 101271813. We acknowledge \textsc{SCAPHYR} for computational facilities at Université de Rennes, France.
\end{acknowledgments}

\bibliography{reference.bib}
%\end{multicols}
\end{document}